Thermal Conductivity due to Spinons in the One-Dimensional Quantum Spin System $Sr_2V_3O_9$

Takayuki Kawamata[1], Masanori Uesaka[1], Mitsuhide Sato[1], Koki Naruse[1], Kazutaka Kudo[2,*], Norio Kobayashi[2], and Yoji Koike[1]

[1]Department of Applied Physics, Tohoku University, 6-6-05 Aoba, Aramaki, Aoba-ku, Sendai 980-8579, Japan

[2]Institute for Materials Research, Tohoku University, 2-1-1 Katahira, Aoba-ku, Sendai 980-8577, Japan

[*]Present address: Department of Physics, Okayama University, 3-1-1 Tsushima-naka, Kita-ku, Okayama 700-8530, Japan

We have measured the thermal conductivity along different directions of the $S = 1/2$ one-dimensional (1D) spin system $Sr_2V_3O_9$ in magnetic fields up to 14 T.   It has been found that the thermal conductivity along the $[10\bar{1}]$ direction, $\kappa_{[10\bar{1}]}$, is large and markedly suppressed by the application of magnetic field, indicating that there is a large contribution of spinons to $\kappa_{[10\bar{1}]}$ and that the spin chains run along the $[10\bar{1}]$ direction.   The maximum value of the thermal conductivity due to spinons is ~14 W/Km along the $[10\bar{1}]$ direction, supporting the empirical law that the magnitude of the thermal conductivity due to spinons is roughly proportional to the antiferromagnetic interaction between the nearest neighboring spins.

## 1. Introduction

In some low-dimensional quantum spin systems, a large contribution of magnetic excitations to the thermal conductivity has been observed.[1,2]  Therefore, these systems are expected to be utilized as highly thermal-conducting materials.  Although the mechanism has not been understood fully, it has empirically been known that large values of the thermal conductivity due to spins, $\kappa_{spin}$, are owing to the high velocity of magnetic excitations, so that the large bandwidth of magnetic excitations is suitable for large $\kappa_{spin}$ and the antiferromagnetic (AF) correlation between the nearest neighboring spins is more suitable than the ferromagnetic one.[3]  In the most understanding one-dimensional (1D) AF Heisenberg spin systems with the spin quantum number $S = 1/2$, it has theoretically been proposed that the thermal conduction due to magnetic excitations, namely, spinons is ballistic.[4-6]  In fact, a large contribution of spinons to the thermal conductivity has been observed in $S = 1/2$ 1D AF Heisenberg spin systems such as $Sr_2CuO_3$ and $SrCuO_2$.[7,8]  Moreover, the ballistic nature of the thermal conduction due to spinons has experimentally been confirmed in $Sr_2CuO_3$.[9,10] In these compounds, the AF interaction between the nearest neighboring spins, $J$, is as large as more than 2000 K,[11-13] so that the bandwidth of spinons is very large.  This leads to the large contribution of spinons to the thermal conductivity, while the thermal conductivity is little affected by the application of magnetic fields up to 14 T.[14]  On the other hand, the thermal conductivity due to spinons, $\kappa_{spinon}$, in systems with small $J$ values has not been investigated so much, though it is guessed to be affected by the application of a magnetic field comparable with $J/g\mu_B$ ($g$: the $g$-factor, $\mu_B$: the Bohr magneton).

The compound $Sr_2V_3O_9$ is regarded as a $S = 1/2$ 1D AF Heisenberg spin system with a small $J$ value of 82 K estimated from the Bonner-Fisher-type temperature-dependence of the magnetic susceptibility in the polycrystalline sample.[15]  $Sr_2V_3O_9$ contains three kinds of vanadium ions in the unit cell.  Two of them are nonmagnetic $V^{5+}$ ions located in $VO_4$ tetrahedra, and the rest is $V^{4+}$ ions with $S = 1/2$ located in $VO_6$ octahedra.    The $VO_6$ octahedra are connected with each other by sharing an oxygen ion at the corner along the [101] direction, as shown in Fig. 1.    Since the ac-plane with a magnetic network is weakly stacked along the $b$-axis, it appears that the spin chains run along the [101] direction.    From both band calculations using the local density approximation,[15] using the extended Hückel tight-binding method[16] and using the first-principles density-functional theory,[17] however, it has been suggested that the magnitude of the so-called super-superexchange interaction mediated by two oxygen ions of a $VO_4$ tetrahedron along the [10$\bar{1}$] direction is much stronger than that of the superexchange interaction along the [101] direction.    The electron spin resonance (ESR) results have also supported the band calculations.[18]    Therefore, the spin chains might run along not the [101] direction but the [10$\bar{1}$] direction.

We have grown large-sized single crystals of $Sr_2V_3O_9$ and measured the thermal conductivity along the [10$\bar{1}$] direction, $\kappa_{[10\bar{1}]}$, along the [101] direction, $\kappa_{[101]}$, and along the $b$-axis, $\kappa_b$, in magnetic fields parallel to the respective heat current.[19]    As a result, we have found marked suppression of the peak at low temperatures only in the temperature dependence of $\kappa_{[10\bar{1}]}$ by the application of magnetic field.    We have concluded that $\kappa_{\text{spinon}}$ is suppressed by the application of magnetic field, because the AF correlation along the spin chains is disturbed by magnetic fields whose values are comparable with that of $J/g\mu_B$ in $Sr_2V_3O_9$.    In our previous study, however, there was a possibility that the suppression was not dependent on the direction of the thermal conductivity but on the direction of the applied magnetic field.    In this paper, in order to find the contribution of spinons to the thermal conductivity and also to clarify whether the spin-chain direction is the [10$\bar{1}$] or [101] direction clearly, we have measured the thermal conductivity along the [10$\bar{1}$], [101] and $b$-axis directions in magnetic fields up to 14 T parallel to the [10$\bar{1}$], [101] and $b$-axis directions.

## 2. Experimental

Large-sized single crystals of $Sr_2V_3O_9$ were grown by the floating-zone (FZ) method. The details are described in our previous paper.[19]   Thermal-conductivity measurements were carried out by the conventional steady-state method.   One side of a rectangular single-crystal was anchored on the heat sink of copper with indium solder.   A chip-resistance was attached as a heater to the opposite side of the single crystal with GE7031 varnish.   The temperature difference across the crystal was measured with two Cernox thermometers (LakeShore Cryotronics, Inc., CX-1050-SD).

## 3. Results and Discussion

Figure 2 shows our previous results of the temperature dependence of $\kappa_{[10\bar{1}]}$, $\kappa_{[101]}$ and $\kappa_b$ of $Sr_2V_3O_9$ in zero field in the temperature range up to 150 K.[19]   It is found that the magnitude of $\kappa_{[10\bar{1}]}$ is larger than those of $\kappa_{[101]}$ and $\kappa_b$.   Moreover, the peak of $\kappa_{[10\bar{1}]}$ is a little broader than those of $\kappa_{[101]}$ and $\kappa_b$, because an additional shoulder appears around 5 K in $\kappa_{[10\bar{1}]}$ as shown in the inset of Fig. 2.   The thermal conductivity of insulating $Sr_2V_3O_9$ is described as the sum of the thermal conductivity due to phonons, $\kappa_{phonon}$, and due to spinons, $\kappa_{spinon}$.   It is known that the anisotropy of $\kappa_{phonon}$ is usually not larger than that of the thermal conductivity due to magnetic excitations in low-dimensional spin systems[1,2] and that the contribution of $\kappa_{spinon}$ markedly appears along the spin-chain direction.   Therefore, these anisotropic properties of the thermal conductivity imply a large contribution of $\kappa_{spinon}$ to $\kappa_{[10\bar{1}]}$.

It has been reported that a magnetic phase transition from the paramagnetic state to the AF ordered state with decreasing temperature occurs at ~ 5 K in $Sr_2V_3O_9$ due to the inter-chain exchange interaction.[15]   However, it is found that neither anomaly is observed in $\kappa_{[10\bar{1}]}$, $\kappa_{[101]}$ nor $\kappa_b$.   It is known that the thermal conductivity shows a sharp dip at the AF transition temperature, $T_N$, in several three-dimensional antiferromagnets,[20-23] while no sharp dip is observed in several low-dimensional spin systems such as $Sr_2CuO_3$[7,8] and $La_2CuO_4$[24] whose values of $J$ are above 1000 K.   There are two origins of the sharp dip of the thermal conductivity at $T_N$.   One is due to strong scattering of heat carries caused by the critical fluctuations at $T_N$.   The other is due to the increase of the mean free path of heat carries just below $T_N$ owing to the marked reduction of scattering caused by the development of the long-rage order.   In any case, the reason for no sharp dip in $\kappa_{[10\bar{1}]}$, $\kappa_{[101]}$ and $\kappa_b$ at $T_N$ may be that the AF correlation along the spin chains has already developed at high temperatures above $T_N$.

Figure 3 displays the temperature dependence of $\kappa_{[10\bar{1}]}$, $\kappa_{[101]}$ and $\kappa_b$ in zero field and in a magnetic field of 14 T parallel to the $[10\bar{1}]$, $[101]$ and $b$-axis directions together with our previous results.[19]   First, it is noted that the magnitude of the thermal conductivity in zero field is a little different sample by sample, even though the heat-current direction is the same.   This may be due to the difference of the degree of the oxidation of samples during the sample setting to the heat sink using indium solder and so on, because the oxidation of $Sr_2V_3O_9$ single crystals is inferred to change $V^{4+}$ ions with $S = 1/2$ to $V^{5+}$ ions with $S = 0$.   That is, both spin defects and local lattice distortions are inferred to be introduced into the crystals through the oxidation, leading to the suppression of $\kappa_{spinon}$ and $\kappa_{phonon}$.   However, it is found that $\kappa_{[10\bar{1}]}$ is suppressed by the application of magnetic field in any directions, while neither $\kappa_{[101]}$ nor $\kappa_b$ is suppressed.   This result indicates that the suppression is dependent on the direction of the thermal conductivity.   Therefore, it is expected that $\kappa_{spinon}$ is suppressed by the application of magnetic field, while $\kappa_{phonon}$ hardly changes.

It is known that the Hamiltonian of a $S = 1/2$ 1D AF Heisenberg spin system in a magnetic field changes to that of an 1D interacting spinless fermion system by the Jordan-Wigner transformation. Since the magnetic field operates as the chemical potential of fermions, the number of fermions increases by the application of magnetic field.   Since the increase of fermions means that the number of spinons increases, the anti-parallel spin alignment in spin chains of $Sr_2V_3O_9$ is disturbed by the application of magnetic field. Therefore, if phonons were strongly scattered by spinons, both $\kappa_{[10\bar{1}]}$, $\kappa_{[101]}$ and $\kappa_b$ would be suppressed by the application of magnetic field. Considering that only $\kappa_{[10\bar{1}]}$ is actually suppressed by the application of magnetic field, however, it appears that the spinons introduced by the application of magnetic field behave as scatterers of the other spinons rather than operate to carry heat additionally, leading to the suppression of $\kappa_{spinon}$. Furthermore, almost neither change in $\kappa_{[101]}$ nor $\kappa_b$ by the application of magnetic field indicates that the contribution of $\kappa_{spinon}$ is negligible along the [101] and $b$-axis directions. Accordingly, it is concluded that the spin chains run along the [$10\bar{1}$] direction, as suggested from the band calculations[15-17] and the ESR results.[18] Here, it is noted that no change of $\kappa_{spinon}$ has been observed by the application of magnetic fields up to 14 T in other $S = 1/2$ 1D AF Heisenberg spin systems such as $Sr_2CuO_3$ and $SrCuO_2$ with $J \sim 2000$ K.[14] The energy of a magnetic field of 14 T is equivalent to that of 10 K and is not far from the magnitude of $J$ of $Sr_2V_3O_9$. Therefore, it is reasonable that $\kappa_{spinon}$ is suppressed by the application of a magnetic field of 14 T in $Sr_2V_3O_9$ in spite of no magnetic effect on $\kappa_{spinon}$ in $Sr_2CuO_3$ and $SrCuO_2$.

In order to see the degree of the suppression quantitatively, the difference between the thermal conductivities along the [$10\bar{1}$] direction in zero field and in a magnetic field of 14 T, $(\kappa_{14T} - \kappa_{0T})/\kappa_{0T}$, is shown in Figs. 3(a')-(c'). It appears that the value of $(\kappa_{14T} - \kappa_{0T})/\kappa_{0T}$ exhibits the minimum around 5 K in any directions of magnetic field, suggesting that $\kappa_{spinon}$ along the [$10\bar{1}$] direction shows the maximum around 5 K in zero field.

Here, we estimate $\kappa_{spinon}$ in $\kappa_{[10\bar{1}]}$, where both contributions of $\kappa_{spinon}$ and $\kappa_{phonon}$ are included. In the temperature dependence of $\kappa_{[10\bar{1}]}$, only one broad peak is observed around 10 K, indicating that both peaks of $\kappa_{spinon}$ and $\kappa_{phonon}$ are almost overlapped. Therefore, it is hard to estimate $\kappa_{spinon}$ and $\kappa_{phonon}$ separately. On the other hand, both $\kappa_{[101]}$ and $\kappa_b$ are due to only $\kappa_{phonon}$. Accordingly, it is possible to estimate $\kappa_{phonon}$ in $\kappa_{[10\bar{1}]}$ using fitting parameters obtained from the fit of $\kappa_{[101]}$ and $\kappa_b$ with the following equation of $\kappa_{\text{phonon}}$ based upon the Debye model.[25]

$$\kappa_{\text{phonon}} = \frac{k_{\text{B}}}{2\pi^2 v_{\text{phonon}}} \left(\frac{k_{\text{B}}}{\hbar}\right)^3 T^3 \int_0^{\Theta_{\text{D}}/T} \frac{x^4 \mathrm{e}^x}{(\mathrm{e}^x - 1)^2} \tau_{\text{phonon}} \mathrm{d}x, \qquad (1)$$

where $x = \hbar\omega/k_{\text{B}}T$, $\omega$ is the phonon angular frequency, $\hbar$ the Planck constant, $k_{\text{B}}$ the Boltzmann constant, $v_{\text{phonon}}$ the phonon velocity, $\theta_{\text{D}}$ the Debye temperature, and $\tau_{\text{phonon}}$ the relaxation time of the phonon scattering.    The $v_{\text{phonon}}$ is calculated as

$$v_{\text{phonon}} = \frac{k_{\text{B}}\Theta_{\text{D}}}{\hbar}(6\pi^2 n)^{-1/3}, \qquad (2)$$

where $n$ is the number density of atoms.    The phonon scattering rate, $\tau_{\text{phonon}}^{-1}$, is assumed to be given by the sum of scattering rates due to various scattering processes as follows,

$$\tau_{\text{phonon}}^{-1} = \frac{v_{\text{phonon}}}{L_{\text{b}}} + D\omega + A\omega^4 + B\omega^2 T \exp\left(-\frac{\Theta_{\text{D}}}{bT}\right), \qquad (3)$$

The first term represents the phonon scattering by boundaries.    In the present thermal-conductivity measurements, $L_{\text{b}}$ is given by the distance between two terminals of temperature on a single crystal.    The second term represents the phonon scattering by lattice distortions; the third, the phonon scattering by point defects; the fourth, the phonon-phonon scattering in the umklapp process.    First, using Eqs. (1)-(3) with fitting parameters of $D$, $A$, $B$ and $b$ and putting $\theta_{\text{D}}$ at 303 K obtained from the specific heat measurements,[15] the temperature dependence of $\kappa_{[101]}$ and $\kappa_b$ in zero field was fitted fairly well, as shown by red solid lines in Fig. 3.    Values of the best-fit parameters are listed in Table I.    Then, $\kappa_{\text{phonon}}$ in $\kappa_{[10\bar{1}]}$ was estimated, taking into account two matters.    One is that $\kappa_{\text{phonon}}$ at high temperatures above ~40 K is independent of the heat-current direction, because this is the case for $\kappa_{[101]}$ and $\kappa_b$ as seen in Fig. 2.[19]    This is because the phonon-phonon scattering in the umklapp process is dominant at high temperatures and does not seem to be affected by the heat-current direction so much. Therefore, the $b$ value was fixed to be 6.5 between 6.2 and 6.7 and the $B$ value was changed between $8.0 \times 10^{-18}$ s/K and $24 \times 10^{-18}$ s/K. The other is that $\kappa_{\text{spinon}}$ is proportional to $T$ at low temperatures of $k_{\text{B}}T << J$.[26] The A and D values are mainly determined by this matter, because the phonon scattering by point defects and lattice distortions strongly affects values of $\kappa_{\text{phonon}}$ at low temperatures. The $\kappa_{\text{phonon}}$ in $\kappa_{[10\bar{1}]}$ was estimated thus and $\kappa_{\text{spinon}}$ in $\kappa_{[10\bar{1}]}$ was obtained by subtracting $\kappa_{\text{phonon}}$ from the data of $\kappa_{[10\bar{1}]}$, as shown in Fig. 4 and the insets. Values of the best-fit parameters of $\kappa_{\text{phonon}}$ in $\kappa_{[10\bar{1}]}$ are also listed in Table I. It is found that every $\kappa_{\text{spinon}}$ exhibits the maximum at 5 K, which is consistent with the above result that the value of $(\kappa_{14\text{T}} - \kappa_{0\text{T}})/\kappa_{0\text{T}}$ exhibits the minimum around 5 K as shown in Figs. 3(a')-(c'). The maximum values of $\kappa_{\text{spinon}}$ are 14 W/Km, 9 W/Km and 7 W/Km for respective samples. Although theses values are dispersive, they support the empirical law that the magnitude of $\kappa_{\text{spinon}}$ is roughly proportional to $J$, as shown in Fig. 5, where maximum values of $\kappa_{\text{spinon}}$ for various $S = 1/2$ 1D AF spin systems are plotted.[7,8,10,27-31] The difference of the maximum value of $\kappa_{\text{spinon}}$ among the respective samples of $Sr_2V_3O_9$ is inferred to be due to the difference of spin defects and local lattice distortions caused by the oxidation of the samples during the sample setting as mentioned above. Moreover, it is found that the maximum values of $\kappa_{\text{phonon}}$ in $\kappa_{[10\bar{1}]}$ estimated thus are in the range of 8–10 W/Km and are smaller than those of $\kappa_{\text{phonon}}$ in $\kappa_{[101]}$ and $\kappa_b$. It is possible that values of $\kappa_{\text{phonon}}$ in $\kappa_{[10\bar{1}]}$ are a little underestimated, namely, values of $\kappa_{\text{spinon}}$ in $\kappa_{[10\bar{1}]}$ are a little overestimated. Taking into account the uncertainty in the estimation of $\kappa_{\text{spinon}}$, in any case, the empirical law of the magnitude of $\kappa_{\text{spinon}}$ versus $J$ holds good.

Next, we estimate the mean free path of spinons, $l_{\text{spinon}}$, using $\kappa_{\text{spinon}}$ estimated thus. The $\kappa_{\text{spinon}}$ is given by

$$\kappa_{\text{spinon}} = C_{\text{spinon}} v_{\text{spinon}} l_{\text{spinon}}, \qquad (4)$$

where $C_{\text{spinon}}$ and $v_{\text{spinon}}$ are the specific heat and velocity of spinons, respectively. These values are estimated based on the des Cloizeaux-Pearson mode in $S = 1/2$ 1D AF Heisenberg spin systems,[32] and finally $\kappa_{\text{spinon}}$ is given by the following equation at low temperatures of $k_{\text{B}}T/J < 0.15$.[8]

$$\kappa_{\text{spinon}} = \frac{2n_{\text{s}}k_{\text{B}}^2 a}{\pi\hbar}T\int_0^{J\pi/2k_{\text{B}}T}\frac{x^2\mathrm{e}^x}{(\mathrm{e}^x+1)^2}l_{\text{spinon}}\mathrm{d}x \tag{5}$$

Here, $x = \varepsilon/k_{\text{B}}T$, $\varepsilon$ is the energy of spinons, $n_{\text{s}}$ the number density of spins, $a$ the distance between the nearest neighboring spins in the chain. Using Eq. (5) and neglecting the $x$ dependence of $l_{\text{spinon}}$, the temperature dependence of $l_{\text{spinon}}$ was estimated as shown in Fig. 6. Values of $l_{\text{spinon}}$ at high temperatures above ~10 K have large errors, because Eq. (5) holds good at low temperatures of $k_{\text{B}}T/J < 0.15$. In any case, it is certain that $l_{\text{spinon}}$ markedly increases with decreasing temperature below ~10 K, reaching ~7000 Å at 2 K in the sample whose maximum value of $\kappa_{\text{spinon}}$ is as large as 14 W/Km. The value of $l_{\text{spinon}}$ at 2 K is different sample by sample, which is guessed to be due to the difference of spin defects and local lattice distortions caused by the oxidation of the samples during the sample setting as mentioned above. It has been reported that the length of mean free path of magnetic excitations including spinons is typically several thousands Å at low temperatures[29]. Therefore, the estimated values of $l_{\text{spinon}}$ and $\kappa_{\text{spinon}}$ are valid, even if the values are the different sample by sample.

4. Conclusions

We have measured $\kappa_{[10\bar{1}]}$, $\kappa_{[101]}$ and $\kappa_b$ of single crystals of the $S = 1/2$ 1D AF Heisenberg spin system $Sr_2V_3O_9$ in magnetic fields up to 14 T. It has been found that $\kappa_{[10\bar{1}]}$ is large and markedly suppressed by the application of magnetic field in any directions, while neither $\kappa_{[101]}$ nor $\kappa_b$ is suppressed. Accordingly, it has been concluded that there is a large contribution of spinons to $\kappa_{[10\bar{1}]}$ and that the spin chains run along the $[10\bar{1}]$ direction, as suggested from the band calculations[15-17] and the ESR results.[18]   The temperature dependence of $\kappa_{[101]}$ and $\kappa_b$ has well been expressed with $\kappa_{phonon}$ based upon the Debye model, while that of $\kappa_{[10\bar{1}]}$ has been expressed with the sum of $\kappa_{spinon}$ and $\kappa_{phonon}$.   The maximum value of $\kappa_{spinon}$ in $\kappa_{[10\bar{1}]}$ has been estimated as ~14 W/Km, supporting the empirical law that the magnitude of $\kappa_{spinon}$ is roughly proportional to $J$.   The value of $l_{spinon}$ has been estimated to markedly increase with decreasing temperature below ~10 K, reaching ~7000 Å at 2 K.

Acknowledgments

   We are grateful to M. Matsuda for the helpful discussion.   The thermal-conductivity measurements in magnetic fields were performed at the High Field Laboratory for Superconducting Materials, Institute for Materials Research, Tohoku University.   This work was supported by a Grant-in-Aid for Scientific Research from the Ministry of Education, Culture, Sports, Science and Technology, Japan.

Materials Research 109 (2005) [in Japanese].

Figure captions

Fig. 1. (Color online) Crystal structure of $Sr_2V_3O_9$. $V^{5+}$ ions with $S = 0$ and $V^{4+}$ ions with $S = 1/2$ are located in $VO_4$ tetrahedra and $VO_6$ octahedra, respectively. $VO_6$ octahedra are connected with each other along the [101] direction by sharing an oxygen ion at the corner and are also connected via a $VO_4$ tetrahedron along the [10$\bar{1}$] direction.

Fig. 2. Temperature dependence of the thermal conductivity of $Sr_2V_3O_9$ along the [10$\bar{1}$] direction, $\kappa_{[10\bar{1}]}$, along the [101] direction, $\kappa_{[101]}$, and along the $b$-axis, $\kappa_b$, in zero field. The inset shows the expanded plots at low temperatures.

Fig. 3. (Color online) Temperature dependence of the thermal conductivity of $Sr_2V_3O_9$ (a)-(c) along the [10$\bar{1}$] direction, $\kappa_{[10\bar{1}]}$, (d)-(f) along the [101] direction, $\kappa_{[101]}$, and (g),(h) along the $b$-axis, $\kappa_b$, in zero field and in a magnetic field of 14 T parallel to the [10$\bar{1}$], [101] and $b$-axis directions. Temperature dependence of the difference between the thermal conductivities along the [10$\bar{1}$] direction in zero field and in a magnetic field of 14 T parallel to (a') the [10$\bar{1}$], (b') the [101] and (c') the $b$-axis directions, $(\kappa_{14T} - \kappa_{0T})/\kappa_{0T}$, is also shown. Red solid lines indicate the best-fit results of $\kappa_{phonon}$ using Eqs. (1)-(3) based on the Debye model.

Fig. 4. Temperature dependence of the thermal conductivity of $Sr_2V_3O_9$ along the [10$\bar{1}$] direction, $\kappa_{[10\bar{1}]}$, (closed circle) in zero field, shown in Fig. 2 (also in Fig. 3(a)), $\kappa_{phonon}$ (solid lines) estimated using the Debye model and $\kappa_{spinon}$ (open circle) estimated as $\kappa_{[10\bar{1}]} - \kappa_{phonon}$. Left and right insets indicate the temperature dependence of $\kappa_{[10\bar{1}]}$ shown in Fig. 3(b) and (c), respectively, and $\kappa_{phonon}$ and $\kappa_{spinon}$ estimated in the same way.

Fig. 5. Dependence of the maximum value of $\kappa_{spinon}$ on the AF interaction between the nearest neighboring spins, $J$, for various $S = 1/2$ 1D AF spin systems, LiCuVO$_4$,[27] KCuF$_3$,[28] BaCu$_2$Si$_2$O$_7$,[29] Sr$_2$CuO$_3$,[7,10] SrCuO$_2$,[8,30,31] and Sr$_2$V$_3$O$_9$.

Fig. 6. Temperature dependence of the mean free path of spinons, $l_{spinon}$, of Sr$_2$V$_3$O$_9$ in zero field. The solid circles, open circles and open diamonds indicate $l_{spinon}$ estimated from the thermal conductivity along the [10$\bar{1}$] direction in zero field, $\kappa_{[10\bar{1}]}$, shown in Fig. 3(a), Fig. 3(b) and Fig. 3(c), respectively.

Table I. Parameters used for the fit of the temperature dependence of $\kappa_{\text{phonon}}$ in $Sr_2V_3O_9$ along the $[101]$, $b$-axis and $[10\bar{1}]$ directions in zero field with Eqs. (1)-(3).

| direction | | $L_b$ ($10^{-3}$ m) | $D$ ($10^{-6}$) | $A$ ($10^{-41}$ s$^3$) | $B$ ($10^{-18}$ s/K) | $b$ |
|---|---|---|---|---|---|---|
| $[101]$ | (Fig.3(d)) | 1.1 | 0.2 | 3.8 | 8.0 | 6.7 |
| $[101]$ | (Fig.3(e)) | 1.8 | 2.0 | 3.7 | 9.0 | 6.2 |
| $[101]$ | (Fig.3(f)) | 1.1 | 84 | 1.1 | 22 | 6.2 |
| $b$-axis | (Fig.3(g)) | 0.90 | 5.0 | 1.6 | 19 | 6.2 |
| $b$-axis | (Fig.3(h)) | 0.90 | 8.5 | 1.6 | 20 | 6.2 |
| $[10\bar{1}]$ | (Fig.3(a)) | 1.9 | 1.9 | 2.2 | 12 | 6.5 |
| $[10\bar{1}]$ | (Fig.3(b)) | 1.3 | 90 | 1.4 | 20 | 6.5 |
| $[10\bar{1}]$ | (Fig.3(c)) | 1.4 | 60 | 1.4 | 17 | 6.5 |

Fig. 1

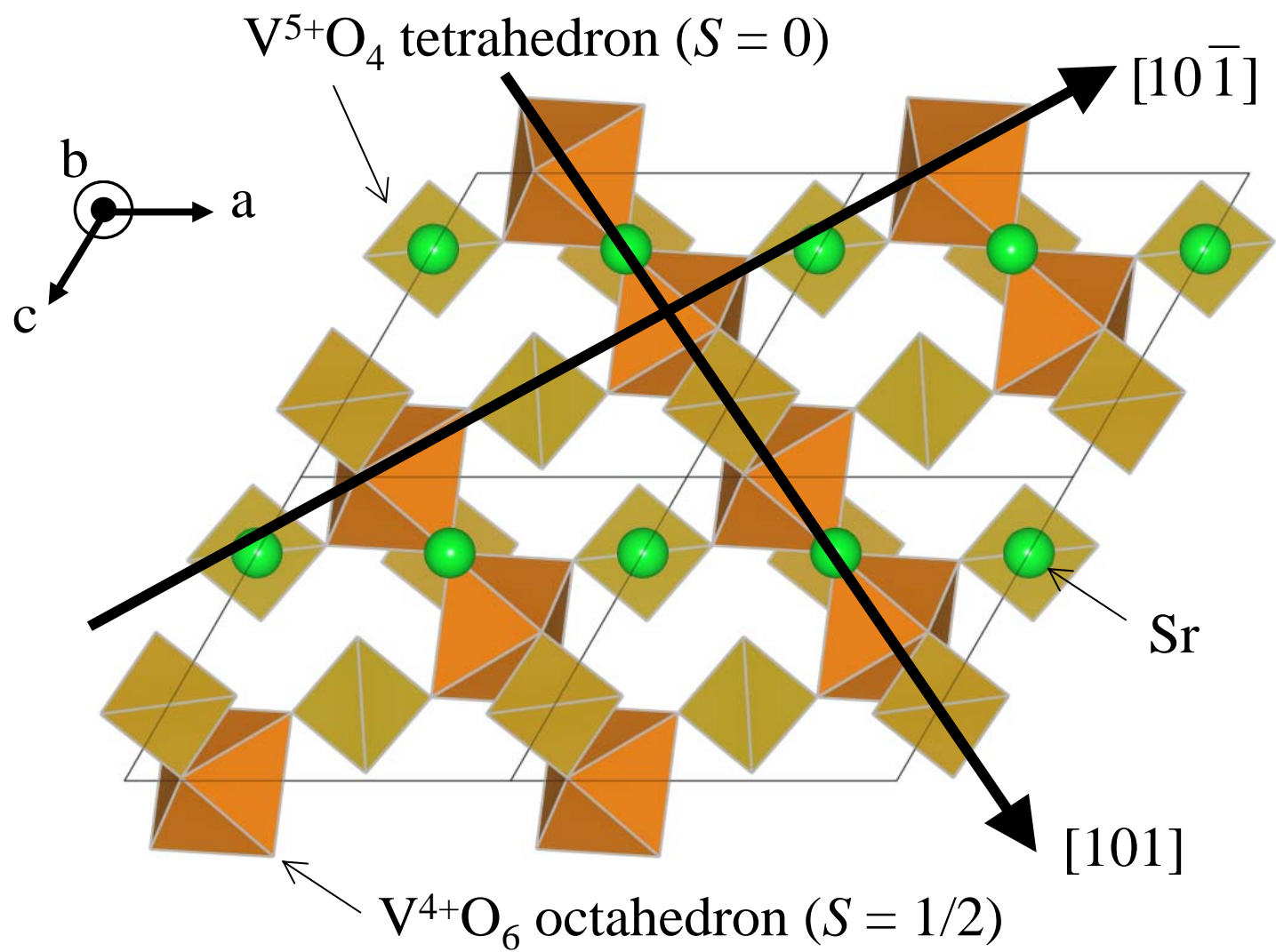

V$^{5+}$O$_4$ tetrahedron ($S = 0$)

[10$\bar{1}$]

Sr

[101]

V$^{4+}$O$_6$ octahedron ($S = 1/2$)

Fig. 2

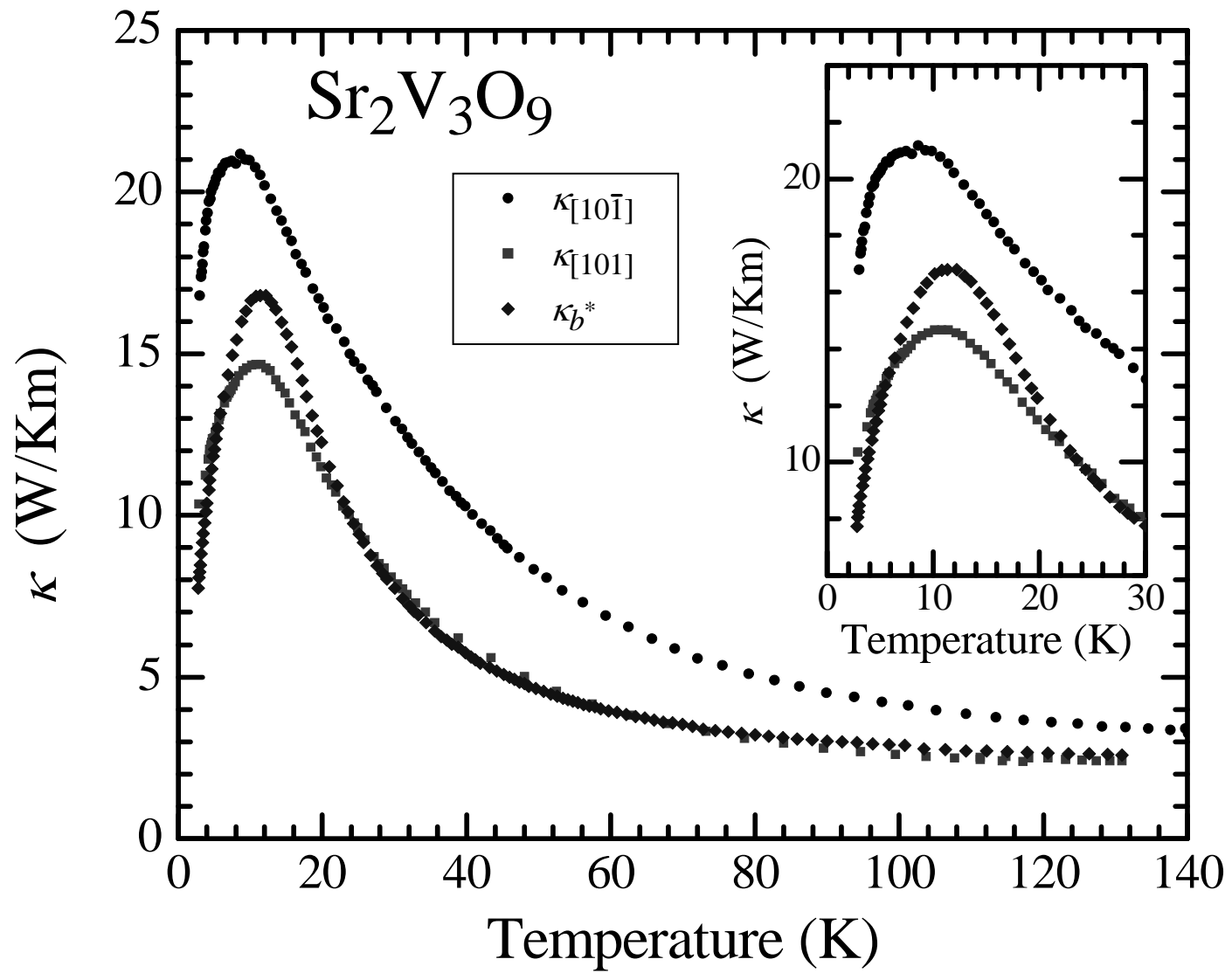

Fig. 3

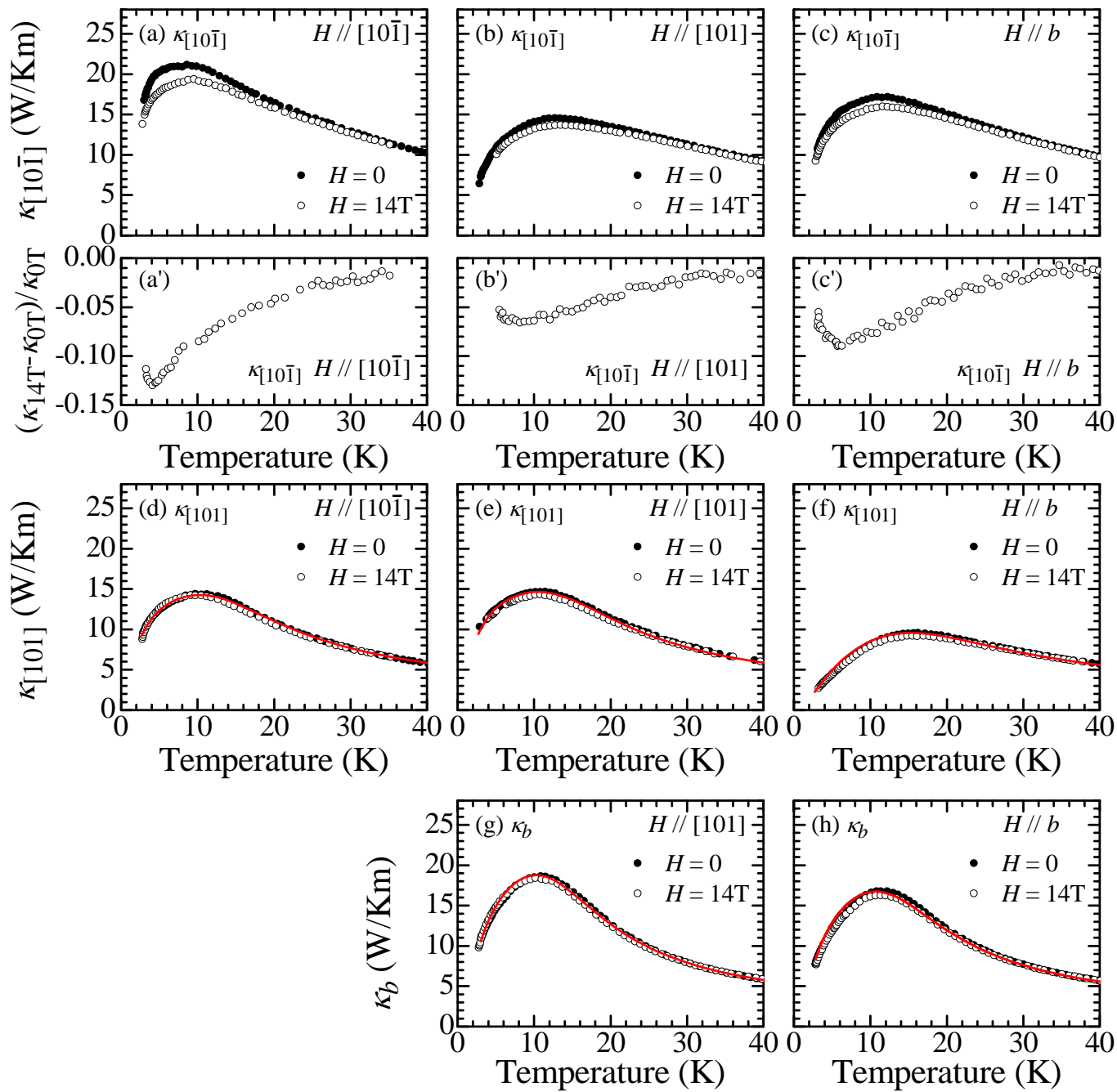

Fig. 4

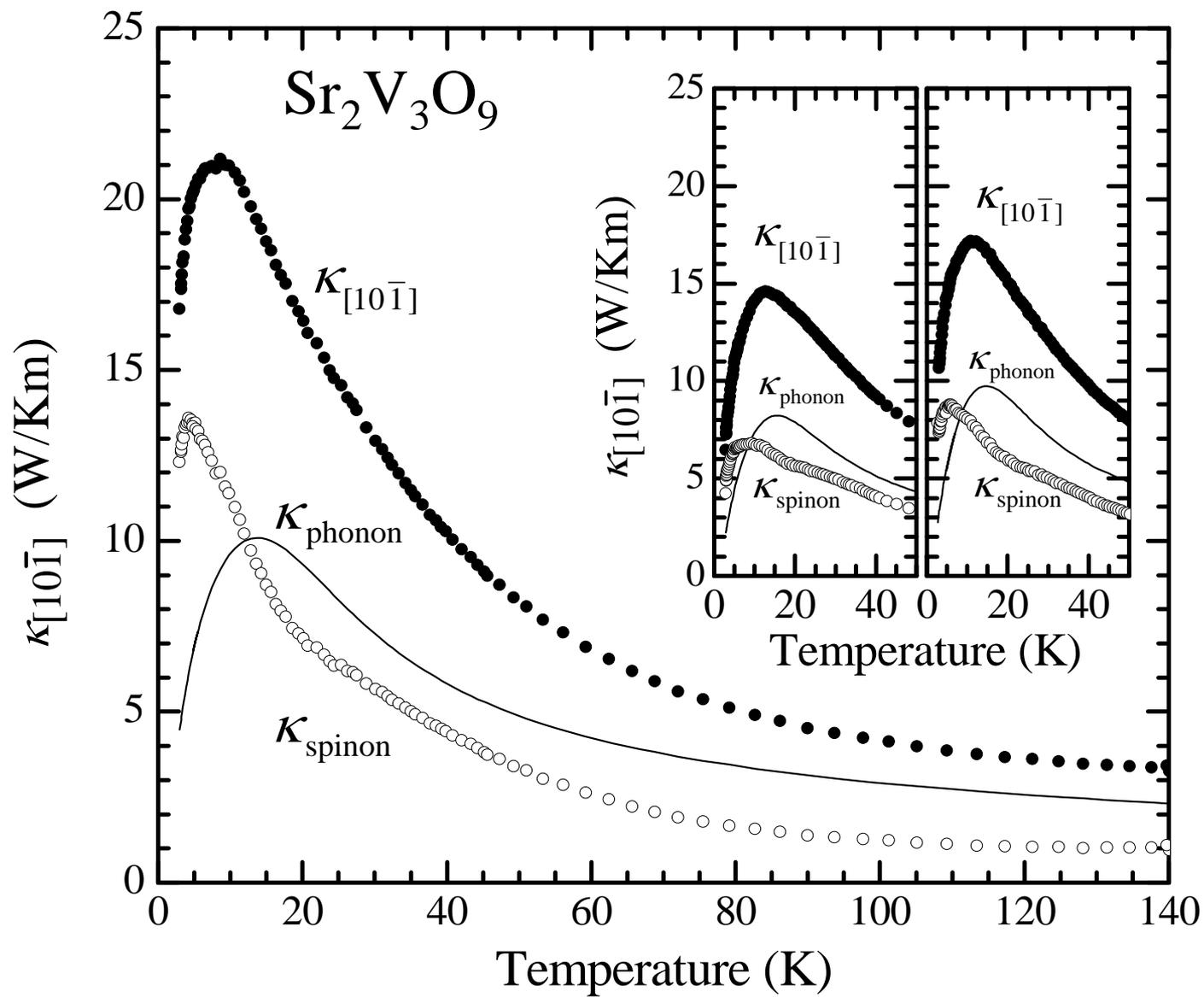

Fig. 5

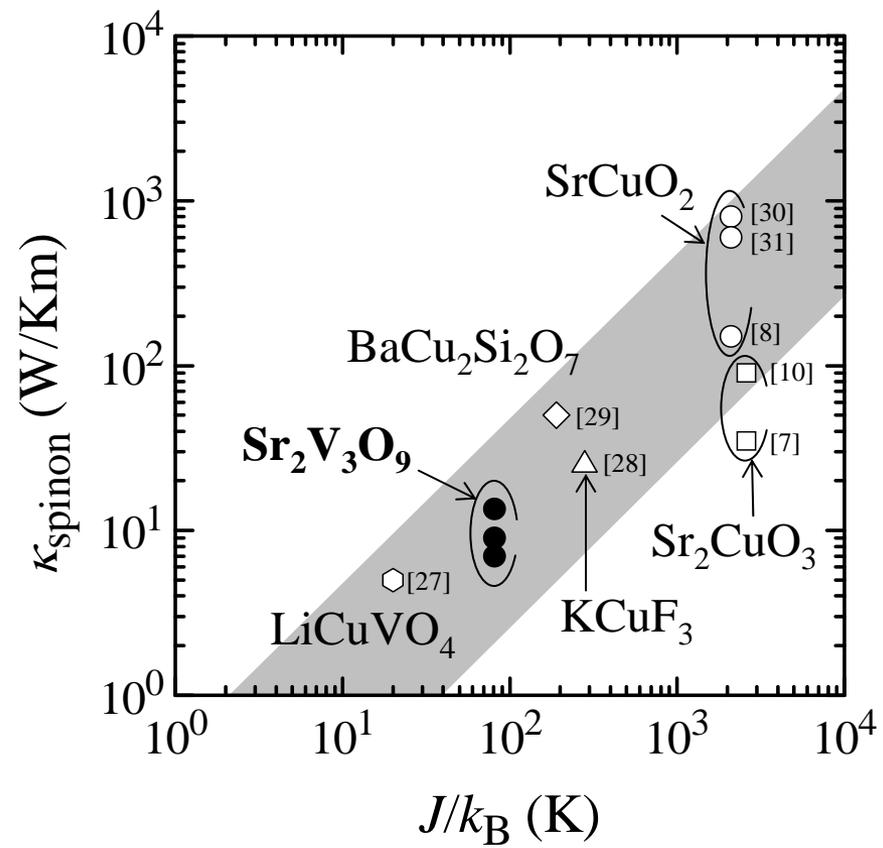

Fig. 6

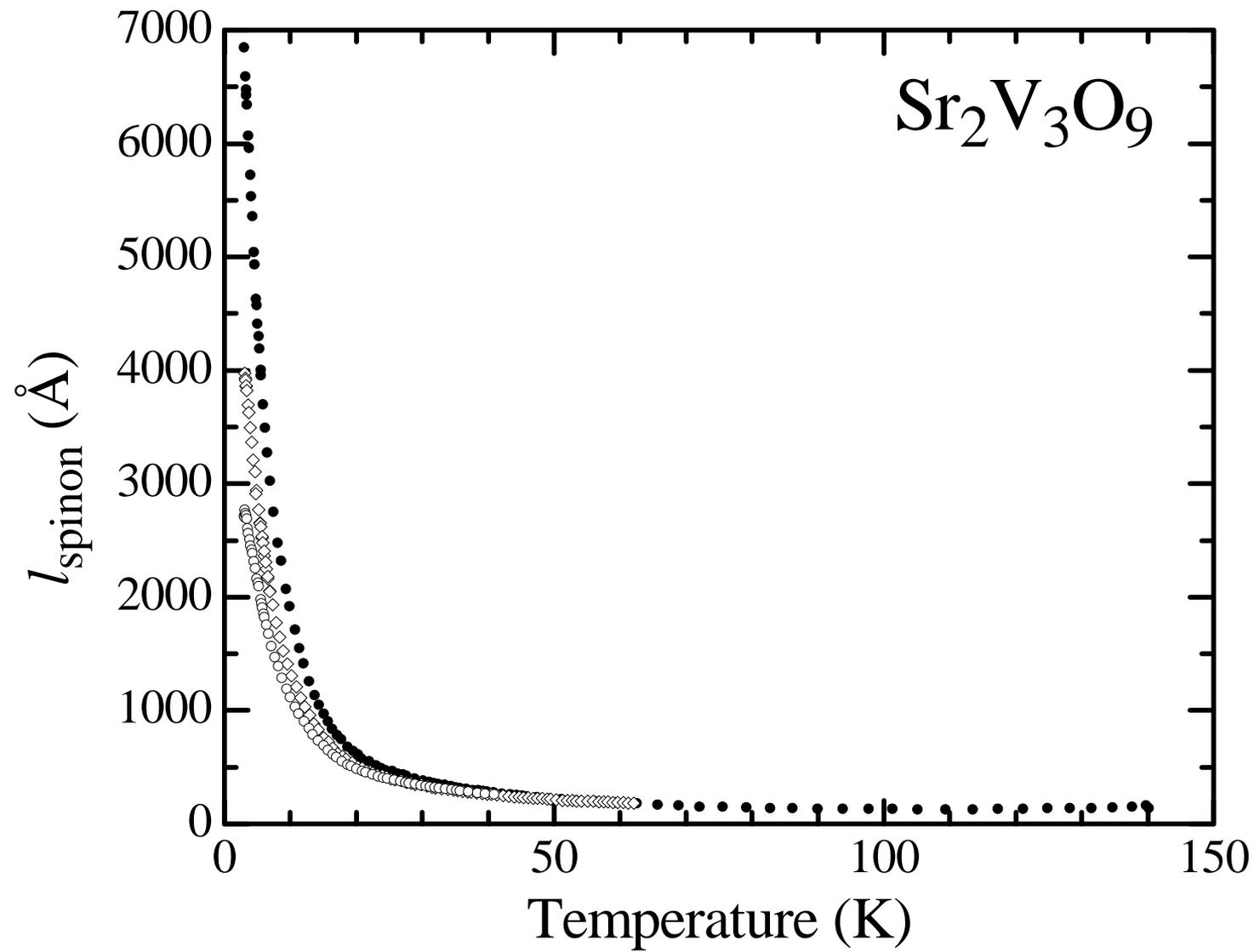